\documentclass[manuscript]{aastex}

\usepackage{natbib}
\usepackage{amsmath}

\begin{document}
\title{Estimation of turbulent diffusivity with direct numerical
simulation of stellar convection}
\author{H. Hotta, Y. Iida and T. Yokoyama}
\affil{Department of Earth and Planetary Science, University of Tokyo,
7-3-1 Hongo, Bunkyo-ku, Tokyo 113-0033, Japan}
\email{ hotta.h@eps.s.u-tokyo.ac.jp}
\begin{abstract}
 We investigate the value of horizontal turbulent diffusivity $\eta$ by numerical
 calculation of thermal convection. In this study, we introduce a new
 method whereby the turbulent diffusivity is estimated by monitoring the time
 development of the passive scalar, which is initially distributed in a
 given Gaussian function with a spatial scale $d_0$.
Our conclusions are as follows:
(1) Assuming
the relation $\eta = L_\mathrm{c}v_\mathrm{rms}/3$
 where $v_\mathrm{rms}$ is the RMS velocity,
the characteristic length $L_\mathrm{c}$ is restricted by the shortest one among the
 pressure (density) scale height and the region depth.
(2) The value of turbulent diffusivity becomes greater with the larger initial
 distribution scale $d_0$.
(3) The approximation of turbulent
diffusion holds better when the ratio of the initial distribution
 scale $d_0$ to the characteristic length $L_\mathrm{c}$ is larger.
\end{abstract}
\section{Introduction}
Turbulent diffusivity has been an important concept for the mean-field
modeling  of the interior convection and dynamo of the Sun and stars
\citep[see the review by][]{2005LRSP....2....1M}.
It is a substantial factor for the
transport of the angular momentum and magnetic field. While the
non-turbulent molecular diffusivities are much smaller, i.e.,
molecular viscosity is $\nu\sim1\ \mathrm{cm^2\ s^{-1}}$ and molecular
magnetic diffusivity is $\eta\sim10^4\ \mathrm{cm^2\ s^{-1}}$ in the
solar convection zone,
the
random advective motion of gases in turbulence is considered to behave
as a strong diffusion. 
The specific value is unknown but
previous studies suggest that the value is around 
$10^{10}$-$10^{13}\ \mathrm{cm^2\ s^{-1}}$
\citep[e.g.][]{1999ApJ...518..508D}.
This value affects predictions of the next solar
maximum. \citep{2007PhRvL..98m1103C,2006ApJ...649..498D,2008ApJ...673..544Y}.
It also affects the symmetry of
the global magnetic field and the strength of the polar field
\citep{2010ApJ...709.1009H,2010ApJ...714L.308H}.
The value determines the difference in rotation speed $\Delta \Omega$
\citep{2011ApJ...740...12H} 
and the propagation speed of the torsional oscillation
\citep{2007ApJ...655..651R}. Thus, the estimation of this value is
crucially important.
\par
Some studies have already estimated the value of turbulent diffusivity on
the solar surface through observations. \cite{1989ApJ...347..529W} investigated the
evolution of active regions and derived the optimum value of turbulent
diffusivity.
\cite{2008ApJ...683.1153C} also estimated the value of turbulent
diffusivity through high resolution observations. They concluded that the
turbulent diffusivity depends on the resolved scale, i.e., the value
becomes smaller with higher resolution.
\cite{2011ApJ...743..133A} also found this type of dependency through
observation of bright points.\par
\cite{2009A&A...500..633K} estimate the value of turbulent diffusion
with numerical simulations of thermal convection using the test field method
\citep{2005AN....326..245S}, which has been adopted for
investigations of many different types of turbulence
\citep{2005AN....326..787B,2008AN....329..725B}. 
A test magnetic field is passively transported by the convection flows
with no back reaction. With the utilization of horizontally averaged
values as a mean
field, the coefficients of the $\alpha$-effect and the turbulent
diffusivity are measured based on the mean-field equations.
\cite{2009A&A...500..633K} report that the value of turbulent
diffusivity is proportional to the square of vertical velocity 
and is approximately
proportional to the wavelength of the test field.
\cite{2011A&A...533A..86C} investigate the value of turbulent
diffusivity with a realistic radiative MHD simulation and estimate the
value of turbulent diffusivity from the decreasing rate of the total
magnetic flux.
\cite{2003A&A...411..321Y} estimate the turbulent magnetic diffusivity
and kinetic viscosity in a forced isotropic turbulence using a similar
way, i.e. from the decay rate of the magnetic field and the velocity field.
\cite{2011SoPh..269....3R} use the cross helicity to estimate the
turbulent magnetic diffusivity in a stratified medium with forced
turbulence.
\cite{2012arXiv1202.1429R} extend this method to numerical calculation
of thermal convection {\bf and} observation of the sun.

\par
In this study, we introduce a new method to estimate the value of
turbulent diffusivity.
We investigate the development of a passive scalar whose
initial condition is the Gaussian function. 
The method is found to be well suited for a Gaussian function at each
time point and
its peak density and spatial extent give us necessary information on
the scalar's kinematics.
A detailed
explanation of the method is given in Section \ref{model}.
The specific aims of this study are:
(1) estimation of turbulent diffusivity of thermal
convection with different sizes of the simulation box;
(2) investigation into the validity of approximation of turbulent diffusion
in thermal convection;
(3) investigation into
the dependence of turbulent diffusivity and the validity of
approximation on the initial distribution scale.

\section{Model}
\subsection{Equations}
The three-dimensional hydrodynamic equation of continuity,
equation of motion, equation of energy, and equation of state are
solved in Cartesian
coordinates $(x,y,z)$, where $x$ and $y$ denote the horizontal
directions and $z$ denotes the vertical direction. The formulations are
almost the same as those used by \cite{2012A&A...539A..30H}.
Equations are expressed as,
\begin{eqnarray}
 && \frac{\partial \rho_1}{\partial t}=-
  \nabla\cdot[(\rho_0+\rho_1){\bf v}]\label{e:e1},\\
 && \frac{\partial {\bf v}}{\partial t}=-({\bf
  v}\cdot\nabla){\bf v}-\frac{\nabla p_1}{\rho_0}
  -\frac{\rho_1}{\rho_0}g{\bf e_z}+\frac{1}{\rho_0}\nabla
 \cdot{\bf \Pi},\label{e:e2}\\
 && \frac{\partial s_1}{\partial t}=-({\bf v}\cdot\nabla)(s_0+s_1)
  +\frac{1}{\rho_0T_0}\nabla\cdot(K\rho_0T_0\nabla s_1)+\frac{\gamma-1}{p_0}({\bf
  \Pi}\cdot\nabla)\cdot{\bf v},\label{e:e3}\\
  && p_1 = p_0 \left(\gamma \frac{\rho_1}{\rho_0}+s_1\right),\label{e:e4}
\end{eqnarray}
where $\rho_0(z)$, $p_0(z)$, $T_0(z)$, and $s_0(z)$ denote the
time-independent, plane-parallel reference density, pressure,
temperature, and entropy, respectively and ${\bf e_z}$
denotes the unit vector along the $z$-direction. $\gamma$ is the ratio
of specific heats, with the value for an ideal gas being $\gamma=5/3$.
$\rho_1$, $p_1$, and $s_1$ denote the fluctuations of density, pressure and
entropy from reference atmosphere, respectively. Note that the entropy
is normalized by specific heat
capacity at constant volume $c_\mathrm{v}$.
The quantity $g$
is the gravitational acceleration, which is assumed to be constant. The
quantity ${\bf \Pi}$ denotes the viscous stress tensor,
\begin{eqnarray}
 \Pi_{ij}=\rho_0\nu
\left[
\frac{\partial v_i}{\partial x_j}+\frac{\partial v_j}{\partial x_i}
-\frac{2}{3}(\nabla\cdot{\bf v})\delta_{ij}
\right],
\end{eqnarray}
and $\nu$ and $K$ denote the viscosity and thermal diffusivity,
respectively. $\nu$ and $K$ are assumed to be constant throughout the
simulation domain.

We assume an adiabatically stratified polytrope for the reference atmosphere
except for entropy:
\begin{eqnarray}
&& \rho_0(z)=\rho_r
\left[
1-\frac{z}{(m+1)H_r}
\right]^m, \\
&& p_0(z)=p_r
\left[
1-\frac{z}{(m+1)H_r}
\right]^{m+1}, \\
&& T_0(z)=T_r
\left[
1-\frac{z}{(m+1)H_r}
\right],\\
&& H_0(z) = \frac{p_0}{\rho_0g},
\end{eqnarray}
where $\rho_\mathrm{r}$, $p_\mathrm{r}$, $T_\mathrm{r}$, and $H_\mathrm{r}$
denote the values of $\rho_0$, $p_0$, $T_0$,
$H_0$ (the pressure scale height) at the bottom boundary $z=0$.
The profile of the reference entropy $s_0(z)$ is defined with a
steady state solution of the thermal diffusion equation
$\nabla\cdot(K\rho_0T_0\nabla s_0)=0$ with constant $K$:
\begin{eqnarray}
&& \frac{ds_0}{dz}=-\frac{\gamma\delta(z)}{H_0(z)},\\
&&\delta(z)=\delta_r\frac{\rho_r}{\rho_0(z)},
\end{eqnarray}
where $\delta$ is the non-dimensional superadiabaticity and
$\delta_\mathrm{r}$ is the value of $\delta$ at $z=0$.
In spite of a non-zero value of superadiabaticity, the adiabatic
stratification is acceptable due to the small value of superadiabaticitiy.
The strength of
the diffusive coefficients $\nu$ and $K$ are expressed with the following
non-dimensional parameters: the Reynolds number
$\mathrm{Re}\equiv v_\mathrm{c}H_\mathrm{r}/\nu$, and the Prandtl number
$\mathrm{Pr}\equiv\nu/K$, where the velocity scale
$v_\mathrm{c}\equiv(8\delta_\mathrm{r}gH_\mathrm{r})^{1/2}$.
In all cases of this study, the parameters are set as $\mathrm{Re}=300$,
$\mathrm{Pr}=1$, $\delta_\mathrm{r}=1\times10^{-2}$. We calculate three
cases with different box sizes (see Table \ref{table_param}). The
horizontal size is the same in all calculations,
i.e. $L_x=L_y=L=26.16H_\mathrm{r}$ and
the number of grids in $x$, $y$ directions are set as $N_x=N_y=1152$.
We adopt three different vertical sizes of box, $L_z=2.18H_\mathrm{r}$,
$1.635H_\mathrm{r}$, and
$1.09H_\mathrm{r}$ for cases 1, 2, and 3 respectively. The number of
grids in these cases are
set as $N_z=96$, $72$, and $48$, respectively. The Rayleigh number, which
is defined as
\begin{eqnarray}
 R_\mathrm{a}\equiv \frac{gH_r^4}{\gamma K \nu}
  \left(
   \frac{\Delta s}{L_z}
  \right),
\end{eqnarray}
in these cases are estimated to be $1.3\times10^5$, $3.4\times10^4$, and
$1.7\times10^4$ respectively, where $\Delta s $ denotes the difference
of entropy between the top and the bottom boundaries.
The calculation domain is $-L/2<x<L/2$, $-L/2<y<L/2$ and $0<z<L_z$.
The boundary conditions and the numerical method are the same as those
used by \cite{2012A&A...539A..30H}. The boundary condition for the $x$, $y$ direction
is periodic for all variables, and the stress free and impenetrative
boundary conditions are adopted and the entropy is fixed, i.e.
$s_1=0$ at $z=0$ and $L_z$.

\subsection{Method for Estimation of Turbulent Diffusivity}\label{model}
In this study, we calculate the evolution of passive scalar to estimate
the value of turbulent diffusivity.
Along with the equations (\ref{e:e1})-(\ref{e:e4}),
we simultaneously solve the advection equation of the passive scalar as
\begin{eqnarray}
 \frac{\partial Q}{\partial t}=-\nabla\cdot(Q{\bf v}),\label{advection}
\end{eqnarray}
where $Q$ is passive scalar density. Although in
eq. (\ref{advection}), the diffusion term does not appear explicitly, we
use tiny artificial viscosity on the passive scalar, a technique which
is adopted in
\citep{2009ApJ...691..640R}.
Its initial condition is set as 
\begin{eqnarray}
 Q(x,y,z,t=0) = \exp\left(-\frac{x^2+y^2}{d_0^2}\right).
  \label{e:initial}
\end{eqnarray}
We adopt three initial conditions, i.e. $d_0=2.5H_\mathrm{r}$,
 $5.0H_\mathrm{r}$, and $7.5H_\mathrm{r}$ for each of the different depth settings
(cases 1-3); hence the total number of cases is nine.
In the initial condition the passive scalar
does not depend on $z$, since we focus on
the turbulent diffusion in the horizontal direction.
Since the transport of the passive scalar is assumed to be approximated by a diffusion
process with constant diffusivity $\eta$, then its density 
should obey the two-dimensional diffusion
equation as
\begin{eqnarray}
 \frac{\partial Q}{\partial t}=\eta
\left(\frac{\partial^2}{\partial x^2}+\frac{\partial^2}{\partial
 y^2}\right)Q.
\label{e:diffusion}
\end{eqnarray}
When the calculation domain is infinite, the analytical solution of
eq. (\ref{e:diffusion}) with the initial condition of
eq. (\ref{e:initial}) is expressed as
\begin{eqnarray}
 Q=\left(\frac{d_0}{d}\right)^2\exp\left(-\frac{x^2+y^2}{d^2}\right),
\end{eqnarray}
where, $d^2=4\eta t+d^2_0$. In this study we adopt
periodic boundary conditions; thus the analytic
solution is given by the periodic superposition of the above formula and
can be expressed as
\begin{eqnarray}
 Q = \sum_{i=-\infty}^{\infty}\sum_{j=-\infty}^{\infty}
  \left(\frac{d_0}{d}\right)^2
  \exp{\left[-\frac{(x-iL)^2+(y-jL)^2}{d^2}\right]}.
\end{eqnarray} 
When the width of the Gaussian function is narrower than box size ($d<L$), the
analytical solution in the range, $-L/2<x<L/2$ and $-L/2<y<L/2$, can be approximated as
\begin{eqnarray}
 Q \sim \sum_{i=-1}^{1}\sum_{j=-1}^{1}
  \left(\frac{d_0}{d}\right)^2
  \exp{\left[-\frac{(x-iL)^2+(y-jL)^2}{d^2}\right]}.
  \label{eq_fit}
\end{eqnarray}
We estimate the value of turbulent diffusivity by the following steps:
\begin{enumerate}
 \item The advection eq. (\ref{advection}) is calculated with
       the obtained velocity of thermal convection.
 \item The obtained passive scalar in each step is vertically averaged as
       \begin{eqnarray}
	\tilde{Q} = \frac{1}{L_z}\int_0^{L_z} Qdz.\label{e:average}
       \end{eqnarray}
       Note that by using this method, we will obtain an averaged turbulent
       diffusivity along the $z$-direction.
 \item The result of averaging, i.e., eq. (\ref{e:average}), is
       fitted with eq. (\ref{eq_fit}). Note that
       the fitting has only one parameter $d(t)$, and this parameter has
       information on both the height and the width of the Gaussian
       function. 
 \item According to the analytical relation, $d^2=4\eta t+d_0^2$, we
       obtain the value of turbulent diffusivity from the slope of $d^2(t)$.
\end{enumerate}

\section{Results \& Discussion}
\begin{figure}[htbp]
 \centering
 \includegraphics[width=15cm]{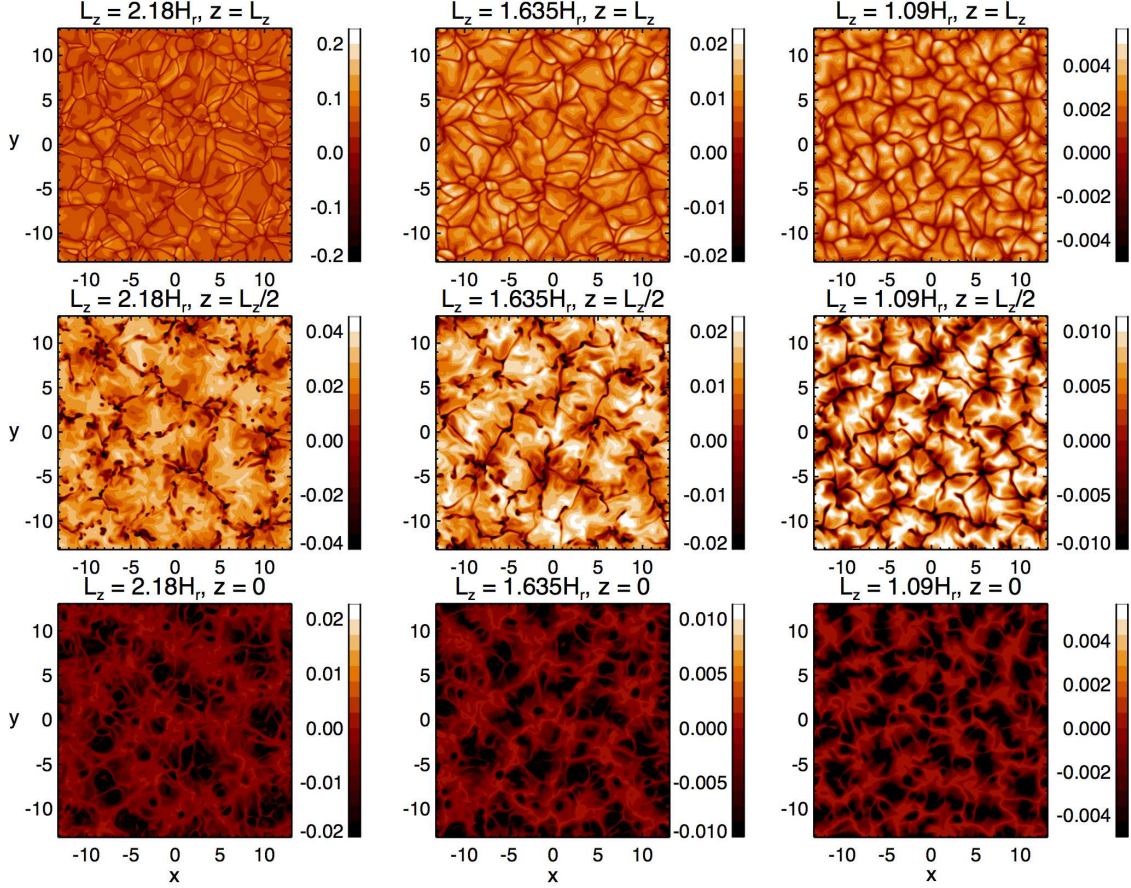}
 \caption{Contour of entropy ($s_1$). The three panels in the left, middle,
 and right columns
 correspond to the results of cases 1, 2, and 3, respectively.
 The rows in the top, middle
 and bottom columns show the plot at $z=L_z$, $L_z/2$, and $0$,
 respectively. \label{conv}}
\end{figure}

Figure \ref{conv}
shows the results of our hydrodynamic calculation. The three panels in
the left,
middle and right columns show the contours of entropy in cases 1, 2 and 3,
respectively. Due to the large Rayleigh number, the velocity with a large
box size is high (see the third row in Table \ref{table_param}).
A detailed
investigation of cell size distribution will be reported in our forthcoming
paper (Iida et al, in preparation).
\par
\begin{figure}[htbp]
 \centering
 \includegraphics[width=15cm]{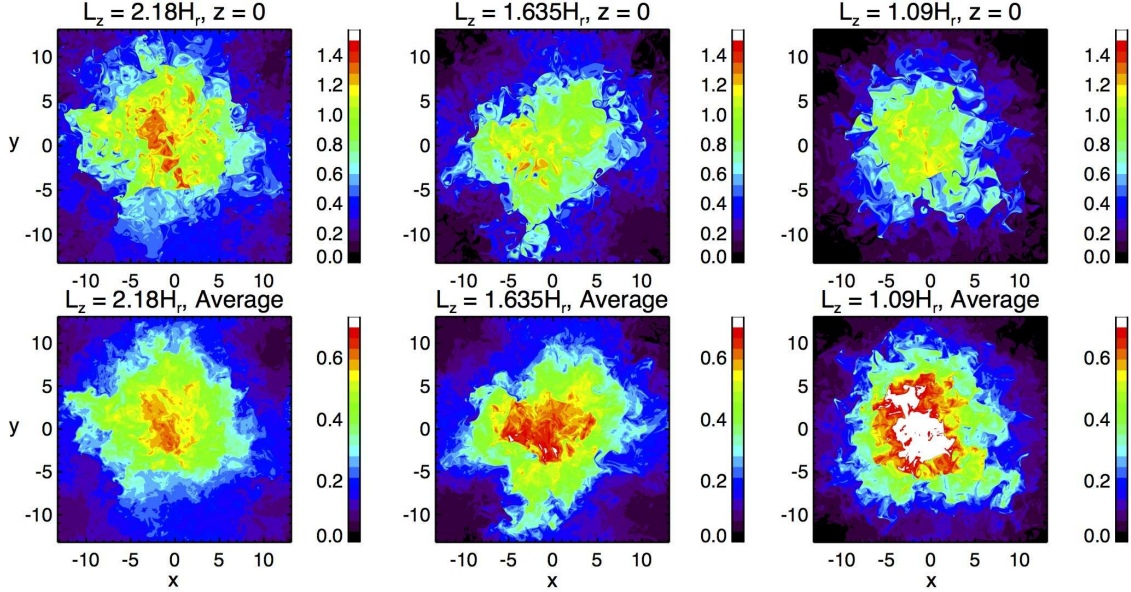}
 \caption{Contour of passive scalar at $t=75H_\mathrm{r}/v_\mathrm{c}$. 
 The panels in the left, middle, and right
 columns show the results in cases 1, 2, and 3, respectively. The panels in
 the top row show the contour of passive scalar $Q$ at $z=0$. 
 The bottom row shows the plot of passive scalar density averaged over $z$,
i.e.,$\tilde{Q}$ defined by eq.
 (\ref{e:average}).\label{passive}}
\end{figure}
\begin{figure}[htbp]
 \centering
 \includegraphics[width=15cm]{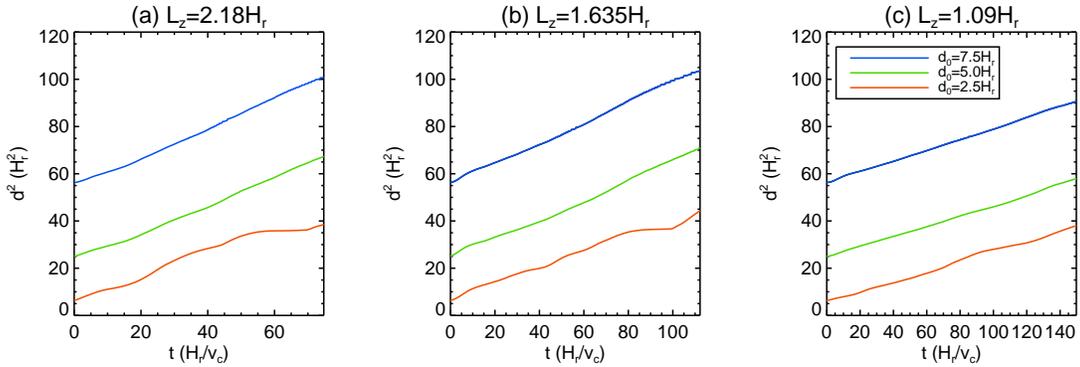}
 \caption{
Distribution range $d^2$ of the passive scalar as functions of time.
 Panels a, b, and c show the dependence of $d^2$
  with the sizes of boxes
 $L_z=2.18H_\mathrm{r}$, $1.635H_\mathrm{r}$, and
 $1.09H_\mathrm{r}$, respectively. 
 The blue, green, and red lines show
 the results with $d_0=7.5H_\mathrm{r}$, $5.0H_\mathrm{r}$, and
 $2.5H_\mathrm{r}$, respectively.
\label{fitting}}
\end{figure}
\begin{figure}[htbp]
 \centering
 \includegraphics[width=15cm]{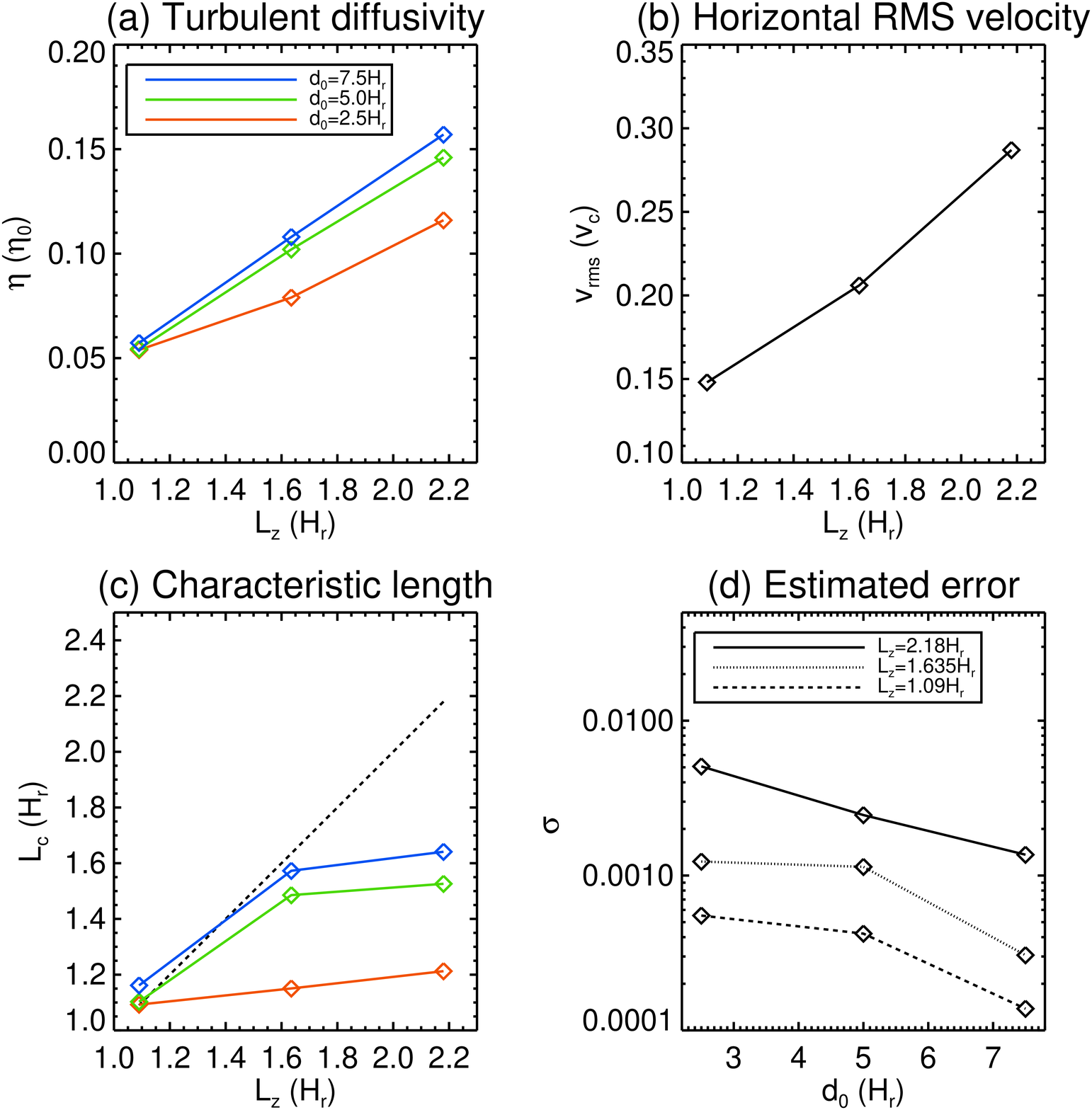}
 \caption{(a) Dependence of turbulent diffusivity on the size of box.
 The value of turbulent diffusivity is normalized by $\eta_0=v_\mathrm{c}H_\mathrm{r}$
 (b) Dependence of horizontal RMS velocity on the size of box. 
 (c) Dependence of characteristic length on size of box. The dashed line shows
 the size of box.
 (d) Dependence of estimated error on the width of the Gaussian function.
 In panels a and b, the blue, green, and red lines show
 the results with $d_0=2.5H_\mathrm{r}$, $5.0H_\mathrm{r}$, and
 $7.5H_\mathrm{r}$, respectively.
In panel d, the solid, dotted and dashed lines show the results with
 $L_z=2.18H_\mathrm{r}$, $1.635H_\mathrm{r}$, and $1.09H_\mathrm{r}$, respectively.
 \label{scaling}}
\end{figure}

Figure \ref{passive} shows the contour of the passive scalar whose
width of the Gaussian function
at the initial condition is $d_0=2.5H_\mathrm{r}$. We can see that the passive scalar
is diffused with turbulent convection. 
The dependences of $d^2$ on $t$ are
provided in Figure \ref{fitting},
and are shown to be almost linear.
This shows the validity of the diffusive description for the turbulent
transport by the convective motion.
The estimated turbulent diffusivity is shown in Figure
\ref{scaling}a.
It is derived through linear fittings to the
curves in Figure \ref{fitting} in the range of $0<t<t_\mathrm{max}$
where $t_\mathrm{max}$ is chosen so as to reduce the
fitting error; it is given in Table \ref{table_param}.
\par
The scaling behavior of the obtained diffusion is studied by changing the
depth of the simulation box in cases 1, 2 and 3.
In Figure \ref{scaling}a, the blue, green, and red lines
show the values of turbulent diffusivity with $d_0=7.5H_\mathrm{r}$,
$5.0H_\mathrm{r}$, and $2.5H_\mathrm{r}$, respectively. 
The value of turbulent diffusivity scales with the size of the box, which is
discussed in the next paragraph. The value of turbulent diffusivity also
scales with the initial width of the Gaussian function $d_0$. 
Using a wider Gaussian
function makes the larger size of the convection cell work more
efficiently and generates
a larger value of turbulent diffusivity.
\par
In the mean field model, it is thought that the coefficient
of turbulent diffusion can be expressed as
$\eta = L_\mathrm{c} v_\mathrm{rms}/3$,
where $L_\mathrm{c}$ is the characteristic length scale of turbulence
and $v_\mathrm{rms}$ is the root-mean-square (RMS) velocity.
The value of turbulent diffusivity is obtained in this study, and we
can estimate the value of $L_\mathrm{c}$ based on the mean field model i.e.,
$L_\mathrm{c}=3\eta/v_\mathrm{rms}$.
The estimated horizontal RMS velocities and characteristic length scale are
shown in Figure \ref{scaling}b and c, respectively.
We discuss the dependence of $L_\mathrm{c}$ on the box size
with $d_0=5.0H_\mathrm{r}$ and $7.5H_\mathrm{r}$.
With a smaller box, i.e. $1.635H_\mathrm{r}$ (case 2) and
$1.09H_\mathrm{r}$ (case 3), the characteristic lengths are almost the same as
the sizes of the boxes (the size of the box is indicated by the dashed line). It
is natural that the largest
cell size is determined by the size of the box and that the largest cell
is most
effective for advecting the passive scalar.
Although we expected that the characteristic length
of case 1 would also be the same as $L_z$, the obtained characteristic
length scale was smaller than $L_z$ even with $d_0=5.0H_\mathrm{r}$ and
$7.5H_\mathrm{r}$.
A possible reason for this result is that the characteristic
length $L_\mathrm{c}$ is
restricted by the convection cell, which is also limited vertically by
the pressure scale height ($H_\mathrm{r}$) or the density scale height
($\gamma H_\mathrm{r}\sim1.67H_\mathrm{r}$). Although $L_\mathrm{c}$ should be
evaluated in the horizontal scale, the
mixture of the passive scalar may occur at approximately the same
distance with the vertical scale.
It should also be noted that when
the narrowest Gaussian function, i.e. $d_0=2.5H_\mathrm{r}$ (red
line), is used, the characteristic lengths are restricted by the width of the
Gaussian function in cases 1 and 2.
\par
Next, we discuss the validity of the approximation of turbulent diffusivity
quantitatively. We calculate the estimated error of the
linear fitting of $d^2$ as:
\begin{eqnarray}
 \sigma =
  \sqrt{\frac{1}{N-2}\sum_{n=1}^N
  \left[
   \frac{d^2(t_n)-\overline{d^2(t_n)}}{\overline{d^2(t_n)}}
  \right]^2},
\end{eqnarray}
where $N$ is the number of data points along the time, $d^2(t_n)$ is the $n$-th
 estimated result and $\overline{d^2(t_n)}$ is the $n$-th result of the
 fitted line. In Figure \ref{scaling}d, we found a dependence of
 $\sigma$ on $d_0$, i.e. $\sigma$ is larger
 with narrower $d_0$.
Although the qualitative relation is not clear,
it indicates that the estimated error $\sigma$ tends to become smaller
 with a larger ratio $d_0/L_\mathrm{c}$ of the initial width of Gaussian function to the
 characteristic length (${\bf L_\mathrm{c}}$)
\section{Summary}
 We investigated the value of horizontal turbulent diffusivity $\eta$ by
 a numerical
 calculation of thermal convection. In this study, we have introduced a new
 method, whereby the turbulent diffusivity is estimated by monitoring the time
 development of the passive scalar, which is initially distributed in a
 given Gaussian function with a spatial scale $d_0$.
Our conclusions are as follows:
(1) Assuming
the relation $\eta = L_\mathrm{c}v_\mathrm{rms}/3$
 where $v_\mathrm{rms}$ is the RMS velocity,
the characteristic length $L_\mathrm{c}$ is restricted by the shortest one among the
 pressure (density) scale height and the region depth.
(2) The value of turbulent diffusivity becomes larger with a larger initial
 distribution scale $d_0$.
(3) The approximation of turbulent
diffusion holds better when the ratio of the initial distribution
 scale $d_0$ to the characteristic length $L_\mathrm{c}$ is larger.
\par
Conclusion (2) is consistent with the results of observational study
\citep{2008ApJ...683.1153C,2011ApJ...743..133A} and a previous
numerical study \citep{2009A&A...500..633K}.
In this study, we do not estimate the correlation length directly from
the thermal convection. This will be achieved in our future work with
an auto-detection technique  and our
 characteristic length ($L_\mathrm{c}$) will be compared with
 directly estimated
 correlation length. We now assume that our characteristic length is an
 average of correlation length at each height (Iida et al. in prepareation).
The turbulent diffusion in the horizontal directions is estimated in this work, but
such estimations are also important in the vertical directions for
addressing the solar dynamo
problem from the viewpoint of the transport of magnetic flux from the surface to
the bottom of the convection zone. 
Such a study will be conducted in the future.\par
 Although turbulent diffusivity averaged in the whole box
 is estimated in this study, the dependence of this estimation on
 the height is important. There are, however, two reasons why it is
 difficult to estimate this dependence
 with our method. First, in our
 calculations the integrated passive scalar density is not conserved at each
 height.
 Second, we found that it is difficult to estimate the diffusivity
 separately for
 each horizontal plane only by solving
 eq. (\ref{advection}) two-dimensionally in the $x-y$ plane the because the
 results show that the passive scalar density is strongly concentrated in the
 boundaries of the convection cells. Such a spatially intermittent
 structure is inappropriate for obtaining a statistical property like the
 turbulent diffusivity. These difficulties will necessitate some
 substantial improvements in our method.
We are also interested in the effect of feedback from the magnetic field to the
convection because of its influence on the turbulent diffusivity 
\citep[e.g.][]{2003A&A...411..321Y,2011SoPh..269....3R}.

\acknowledgements
 The authors thank N. Kitagawa for helpful discussions.
 Numerical computations were, in part, carried out on a Cray XT4 at the Center
 for Computational Astrophysics, CfCA, of the National Astronomical
 Observatory of Japan.
 The page charge for this paper is subsidized by CfCA.
 This work was supported by Grant-in-Aid for JSPS Fellows.
 We have greatly benefited from the proofreading/editing assistance from
 the GCOE program.

\clearpage
\begin{table}
\begin{center}
\caption{Parameters of study's calculation.\label{table_param}}
\begin{tabular}{cccc}
\hline\hline
Case & 1 & 2 & 3 \\
\hline
$L_x\times L_y\times L_z$ $(H_\mathrm{r}^3)$
 & $26.16^2\times2.18$ & $26.16^2\times1.635$ & $26.16^2\times1.09$ \\
 $N_x\times N_y\times N_z$ & $1152^2\times96$ & $1152^2\times72$ & $1152^2\times48$ \\
 $\rho_0(L_z)/\rho_0(0) $ & 22 & 4.9 & 2.4 \\
$v_\mathrm{rms}$ $(v_\mathrm{c})$ 
 & 0.287 & 0.206 & 0.148 \\
 $t_\mathrm{max}$ $(H_\mathrm{r}/v_\mathrm{c})$
 & 75 & 112.5 &150  \\
\hline
\end{tabular}
\end{center}
\end{table}

\end{document}